# Magnetic excitations in infinite-layer nickelates

H. Lu[1,3], M. Rossi[1], A. Nag[2], M. Osada[3], D. F. Li[1], K. Lee[3], B. Y. Wang[3], M. Garcia-Fernandez[2], S. Agrestini[2], Z. X. Shen[1,3], E. M. Been[1], B. Moritz[1], T. P. Devereaux[1,3,4], J. Zaanen[5], H. Y. Hwang[1,3], Ke-Jin Zhou[2]*, W. S. Lee[1]*

**Affiliations:**

[1]Stanford Institute for Materials and Energy Sciences, SLAC National Accelerator Laboratory and Stanford University, 2575 Sand Hill Road, Menlo Park, CA 94025, USA.

[2]Diamond Light Source, Harwell Campus, Didcot OX11 0DE, United Kingdom

[3]Geballe Laboratory for Advanced Materials, Departments of Physics and Applied Physics, Stanford University, Stanford, CA 94305, USA.

[4]Department of Materials Science & Engineering, Stanford University, Stanford, CA 94305, USA.

[5]Instituut-Lorentz for theoretical Physics, Leiden University, Niels Bohrweg 2, 2333 CA Leiden, The Netherlands.

*Correspondence to kejin.zhou@diamond.ac.uk and leews@stanford.edu

**Abstract:**

The discovery of superconductivity in infinite-layer nickelates brings us tantalizingly close to a new material class that mirrors the cuprate superconductors. Here, we report on magnetic excitations in these nickelates, measured using resonant inelastic x-ray scattering (RIXS) at the Ni $L_3$-edge, to shed light on the material complexity and microscopic physics. Undoped $NdNiO_2$ possesses a branch of dispersive excitations with a bandwidth of approximately 200 meV, reminiscent of strongly-coupled, antiferromagnetically aligned spins on a square lattice, despite a lack of evidence for long range magnetic order. The significant damping of these modes indicates the importance of coupling to rare-earth itinerant electrons. Upon doping, the spectral weight and energy decrease slightly, while the modes become overdamped. Our results highlight the role of Mottness in infinite-layer nickelates.

**One Sentence Summary:**
Magnetic excitations in infinite-layer nickelates were revealed for the first time.

**Main Text:**

The mechanism of unconventional, high-temperature superconductivity, as embodied in families of copper oxides, or cuprates, remains a highly controversial subject in condensed matter physics. However, this represents only the tip of a much deeper mystery into the origins of the strange metal phase, the implications of intertwined orders, and the seemingly unending complexity of the cuprate phase diagram, pushing the limit of established mathematical theory. Soon after their discovery, the late P.W. Anderson deserves credit for recognizing that this strangeness can be traced to the strong local electron repulsion (Hubbard $U$) and the peculiar properties of doped Mott insulators, *i.e.* the Mottness, which remain a challenge for the renormalization to more conventional Fermi-liquid or BCS physics (*1, 2*) in large U limit. While large-scale classical simulations of Hubbard-type models have acquired benchmarking status and will be used to validate the first generation of quantum computers (*3*), the case has developed, aided by string-theoretical methods, that the strangeness of cuprate electrons originates in dense, many-body quantum entanglement (*4*).

Understanding the full impact of this many-body entanglement has been made more difficult by the fact that the cuprates were unique. Although many cuprate families exist, it has proven very hard to find other material classes, based on different transition metals, exhibiting similar Mottness, let alone the quasi-two-dimensional structure and small quantum spins deemed to be essential ingredients. The landscape changed very recently with the discovery of superconductivity in doped monovalent infinite-layer nickelates (*5, 6*). Their crystal structures are very similar to the cuprates, with $NiO_2$ planes separated by spacer layers that contain a minimal set of chemical elements- simpler than most of the cuprates. They were predicted (*7-15*) to be isoelectronic to the cuprates: monovalent $Ni^+$ characterized by the same $3d^9$ state as $Cu^{2+}$ in the cuprates.

So, are the nickelates really cuprate cousins in the most important way – Mottness? There exists considerable uncertainty regarding the local Coulomb repulsion, as the atomic *d*-orbitals of monovalent $Ni^+$ are more extended than those of divalent $Cu^{2+}$, a factor that can have a large influence on the magnitude of *U*. The magnetic structure can provide a constraint that allows for proper categorization. In a weakly interacting metal, the magnetic excitations spread over the full bandwidth ($W \sim 3$ eV), while in a spin-density-wave-like system ($U \ll W$) the excitations accumulate at low energies, but exclusively near the ordering wave vectors. At higher energies (near the magnetic zone boundaries) these "paramagnons" should be completely overdamped, decaying in the metallic continuum (*i.e.* Landau damping). However, in a Mott insulator ($U \gtrsim W$) the metallic continuum is "pushed up" by $U$, and instead one is dealing with the excitations of a pure spin system, characterized by long-lived propagating spin waves that survive all the way up to the zone boundary.

Here, we report the first measurements of the spin excitation spectra in infinite-layer nickelates using resonant inelastic X-ray scattering (RIXS). The data leave little doubt that $U$ exceeds the bandwidth and that the infinite-layer nickelates are close cousins of cuprates, although different in one regard. As anticipated by band structure calculations (*8,9,11-15*), and confirmed experimentally (*11*), the rare-earth spacer layers are no longer inert, but instead contribute small metallic Fermi-surface pockets. These pockets influence the spin wave spectrum in the form of extra damping that is absent in the cuprates. The presence of the itinerant rare-earth pocket enriches these nickelates, which provide an intriguing, yet largely unexplored system.

We study ~10 nm films of $Nd_{1-x}Sr_x NiO_2$ grown on a $SrTiO_3$ substrate with a $SrTiO_3$ capping layer of a few unit cells to maintain and orient the nickelate crystalline structure (*5, 16*) (Fig. S1). Figure. 1A (upper) shows the Ni $L_3$-edge X-ray absorption (XAS) spectra of $NdNiO_2$, which exhibits a single peak that corresponds primarily to the $2p^63d^9 \rightarrow 2p^63d^{10}$ transition (*11*), as in cuprates. For RIXS measurement, we adopted an experimental geometry well established to measure magnetic spin excitations in transition metal oxides (Supplementary Materials, Refs. *17-23*), including cuprates. A hierarchy of excitations can be resolved in a RIXS intensity map acquired by tuning the incident photon energy across the Ni $L_3$-edge (Fig. 1B): a fluorescence feature at 3 eV and above, dipole-forbidden *dd*-excitations from 1.2 to 3.0 eV, a peak at ~0.7 eV due to hybridization between Ni *3d* and Nd *5d* orbitals (*11*), and a low energy feature at ~ 0.2 eV whose intensity is maximal near the peak of XAS (Fig. 1A, lower) and is reminiscent of the magnetic excitations seen in RIXS maps of cuprates taken at the Cu $L_3$-edge (*e.g.* Ref. *17-19*). At lower energy, phonon excitations at ~ 0.07 eV can also be resolved (Fig. 1B).

To characterize the behavior of these low energy excitations, we measured detailed momentum-resolved RIXS maps (see Supplementary Materials). As shown in Figs. 2A and 2B, the excitations bear a striking resemblance to spin-1/2 antiferromagnet (AFM) magnons on a square lattice. Namely, they disperse strongly with maxima at (0.5, 0) and (0.25, 0.25), soften towards the conventional AFM ordering wave vector (0.5, 0.5), and exhibit spectral intensity suppression near (0.5, 0) (*24*). Note that the magnetic excitations do not exhibit obvious dispersion along the *c*-axis (Fig. 2C), indicating that they are quasi-two-dimensional. We fit the spectra to a damped harmonic oscillator (DHO) (*19*), which closely resembles an anti-symmetrized Lorentzian (*18,19*) (See Supplementary Materials and Fig. S2), given by

$$\chi''(q,\omega) = \frac{\gamma_q \omega}{(\omega^2 - \varepsilon_q^2)^2 + 4\gamma_q^2 \omega^2} \tag{1}$$

where $\varepsilon_q$ is the undamped mode energy and $\gamma_q$ is the damping factor. Figure 3 shows the fitted $\varepsilon_q$ and $\gamma_q$ along the three high symmetry directions, except for the data near (0, 0) where the fitting is un-reliable where the mode merges with the phonon and the tail of elastic peak. Note that the dispersion extracted from the DHO function is similar to those extracted by anti-Lorentizan function, which essentially tracks the peak position of the spectrum (Supplementary Fig. S3). Interestingly, we also find a significant dispersion of approximately 50 meV along the AFM zone boundary ((*h, 0.5-h*) direction in Fig. 2A and 3A), indicative of substantial exchange interactions beyond nearest-neighbor Ni (*24*). We fit the extracted dispersion to a linear spin wave form for the spin-1/2 square-lattice Heisenberg AFM (*25*), including nearest- and next-nearest-neighbor exchange couplings,

$$H = J_1 \sum_{\langle i,j \rangle} S_i \cdot S_j + J_2 \sum_{\langle i,i' \rangle} S_i \cdot S_{i'} \tag{2}$$

where $S_i$, $S_j$, and $S_{i'}$ denote Heisenberg spins at site $i$, nearest-neighbor sites $j$, and the next-nearest-neighbor sites $i'$, respectively (Supplementary Materials). We find a substantial nearest-neighbor coupling $J_1 = 64.1 \pm 3.4$ meV, and a sizable next-nearest-neighbor coupling $J_2 = -10.2 \pm 2.4$ meV.

The magnon bandwidth (~ 0.2 eV), which measures the strength of the exchange interaction, is quite comparable to that found in the parent cuprates (~ 0.3 – 0.4 eV) and notably higher than other nickelates (*21,22*). While these results provide clarity to the debate in the theoretical literature on the strength of the exchange coupling (*10, 13-15*), the real surprise is the magnitude of the next-nearest-neighbor coupling $J_2$. In the sense of the Zaanen-Sawatzky-Allen classification, one expects the spin exchange interaction to be rather short ranged, as the nickelates are not charge-transfer compounds (*11*) in which the oxygen ligands serve as a major pathway for the super-exchange interaction (*26*). Thus, this indicates a possible long-range RKKY metallic exchange mediated by the Nd $5d$ pockets. While the negative $J_2$ (*i.e.* un-frustrated) should stabilize conventional Néel order, neutron powder diffraction studies indicate a lack of long-range magnetic order (*27*). The results may be obscured by the quality of these metastable materials; however, establishing the fate of the magnetic order is a matter of highest priority as the absence of order would signal a unique form of quantum magnetism.

The other novelty revealed by the fits is in the spin wave damping $\gamma_q$ (Fig. 3). The width of the magnetic peaks (~$2\gamma_q$ of ~130 meV) exceeds the value of our instrument resolution (~37 meV). Interestingly, this width is independent of the mode energy and momentum. The fingerprint of itinerant magnetism would be a damping that increases sharply moving away from the magnetic ordering wave-vector (presumably at (0.5, 0.5)), and due to spin-wave interactions, the damping should be maximal at the magnetic zone boundary. The rather constant $\gamma_q$ suggests that the magnons instead couple to a "heat bath", capable of dissipating only small momenta and energy. This may be consistent with the metallic Nd pockets and their small Fermi-surfaces, as determined from either tight binding models (*11, 15*) or larger experimental estimates (*5*). Incomplete chemical reduction and disorder that may suppress long-range antiferromagnetism also can play a role in damping these excitations.

Let us now discuss the doping evolution of the magnetic excitations in $Nd_{1-x}Sr_xNiO_2$ from x = 0 to 0.225, across the superconducting phase boundary (*28*). As shown Fig. 4A, the magnetic excitations appear to soften with increasing doping concentration and overlap significantly with phonon excitations towards small momentum (see also Fig. S4A). For those momentum positions, we assume the phonons to be doping independent and deduce magnetic spectra from our fitting analysis, as shown in Fig. 4B (see Supplementary Materials), which contains information about the imaginary part of the dynamical spin susceptibility. As a function of momentum, the magnetic spectra are less dispersive (Fig. 4C and Supplementary Fig. S4B) than those in $NdNiO_2$. The magnetic spectral weight (Fig. 4D) decreases gently, consistent with spin dilution as expected in a doped Mott insulator as some spins are replaced by holes. Consequently, the magnetic modes should soften, and the reduced lifetimes would indicate overdamped, relaxational dynamics. To extract the mode energy $\varepsilon_q$ and damping $\gamma_q$, we fit the data to the DHO function (Eq.1 and see Supplementary Materials). Indeed, the superconducting compound acts as a doped Mott-insulator, with $\varepsilon_q$ mildly softened compared to the magnons in the parent compound with a similar dispersion (Fig. 4E). The most significant change is in the $\gamma_q$ along *hh*-direction which increases dramatically compared to the parent nickelate (Fig. 4F), causing substantial asymmetry in the magnetic spectrum (*e.g.* Fig. 4B). Since $\gamma_q \gtrsim \varepsilon_q$, these high energy spin excitations in the superconductor are on the verge of becoming overdamped. Note that the DHO fitting is consistent with spectral moment analysis revealing that the spectra deliver information regarding the short time scales governed by Mottness (see Supplementary Fig. S5).

Compared to cuprates, with quite similar overdamped magnetic modes upon doping, the infinite-layer nickelates show subtle, but significant differences in the evolution of spectral weight and mode energies. In the cuprates, the spectral weight is found to be essentially unchanged and the mode energy increases upon significant doping (*19*). These effects seem, at first sight, counterintuitive, but are rooted in subtle interference effects associated with longer-range hopping, three-site exchange interactions, and the microscopic nature of doped holes (*29,30*). In fact, the doped nickelates exhibit mild softening and loss of spectral weight that conforms more closely to expectations of the simple t-J model (*31*). As charge-transfer insulators, the doped holes in cuprates reside mostly on oxygen, implying relatively extended hole-wavefunctions, which promote the non-local interactions giving rise to the interference effects. The nickelates appear to be more Mott-Hubbard-like (*11, 15*), implying more localized Ni *d*-like character to the hole-wavefunctions and concomitant suppression of non-local interactions. Our results also highlight the lack of a simple, empirical scaling between the superconducting transition temperature $T_c$ and magnetic energy, implying the need to optimize various degrees of freedom for higher $T_c$. More importantly, with the added complexity due to the presence of the Nd metallic pockets, the upcoming challenge is to find out how these microscopic differences impact collective phenomena in Mott systems. This should give further insight into the relationship between the microscopic ingredients determined by chemistry and the mechanism of superconductivity in these many-body-entangled quantum materials.

**Acknowledgments:**

This work is supported by the U.S. Department of Energy (DOE), Office of Science, Basic Energy Sciences, Materials Sciences and Engineering Division, under contract DE-AC02-76SF00515. We acknowledge the Gordon and Betty Moore Foundation's Emergent Phenomena in Quantum Systems Initiative through grant number GBMF4415 for synthesis equipment. We acknowledge Diamond Light Source for providing the beam time at the I21-RIXS beamline under Proposal NT25165.

**Author contributions:** W.S.L. and K.J.Z conceived the research and designed experiment. H.L., M.R., A.N., M.G.-F., S.A., K.J.Z., and W.S.L. conducted the experiment at Diamond Light Source. H.L., M.R., A.N., K.J.Z., and W.S.L. analyzed the data. M.O., D.L., K.L., B.Y.W., and H.Y.H. synthesized and characterized samples for the experiments. W.S.L., K.J.Z., H.L., M.R., A.N., D.L., H.Y.H., E.M.B. B.M., Z.X.S., T.P.D., and J.Z. discussed and interpreted the results. H.L. and W.S.L. wrote the manuscript with input from all authors.

**Competing interests:** Authors declare no competing interests.

**Data and materials availability:** The datasets generated during and/or analyzed during the current study are available from the corresponding author on reasonable request.

**Supplementary Materials:**
Materials and Methods
Figures S1-S5
References (*32-33*)

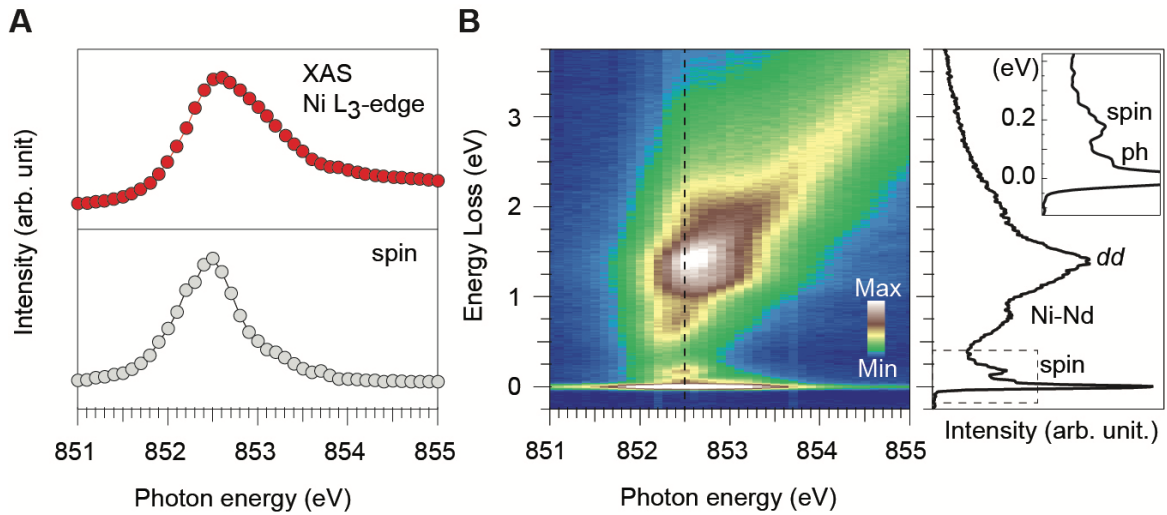

**Fig. 1. XAS and RIXS map of elementary excitations in NdNiO$_2$** (**A**) (upper) Ni $L_3$-edge XAS measured by total electron yield at 15 K in normal incidence geometry. (lower) Resonant profile of spin/magnetic excitations obtained by integrating the RIXS intensity between 0.1 and 0.26 eV. (**B**) RIXS intensity map versus energy loss and incident photon energy across the Ni $L_3$-edge. RIXS spectrum taken at a photon energy of 852.5 eV (black dashed line in B) is shown on the right panel. Inset shows an enlarged view of the spectra within the dashed box. "*dd*", "Ni-Nd", "spin", and "ph" denote orbital excitations within the Ni 3*d* orbitals, spectral feature of Ni 3*d* and Nd 5*d* orbital hybridization, magnetic excitation (~0.2 eV), and phonon (~ 0.07 eV), respectively.

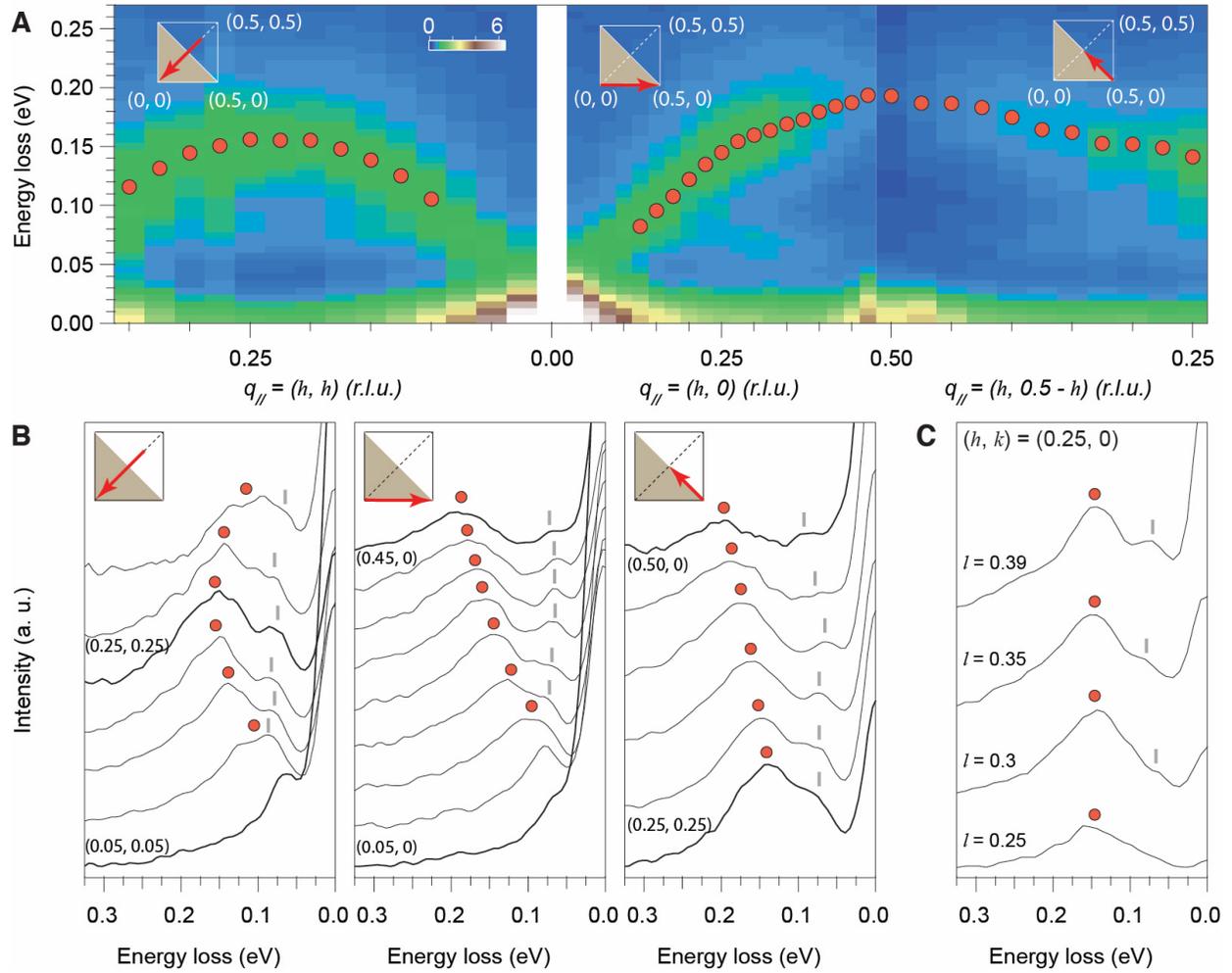

**Fig. 2. Momentum dependent RIXS intensity maps of NdNiO$_2$.** (**A**) RIXS intensity maps versus energy loss and projected in-plane momentum transfer along three high-symmetry directions, as indicated by red arrows in the insets which show a Brillouin zone with the first AFM zone shaded. The red circles indicate peak positions of the magnetic excitation spectra. (**B**) Raw RIXS spectra at representative projected in-plane momentum transfers. The red circles indicate the peak positions of magnetic excitations, while the grey ticks indicate phonon excitations. (**C**) Raw RIXS spectra measured at a fixed in-plane momentum (0.25, 0) with different out-of-plane momentum $l$.

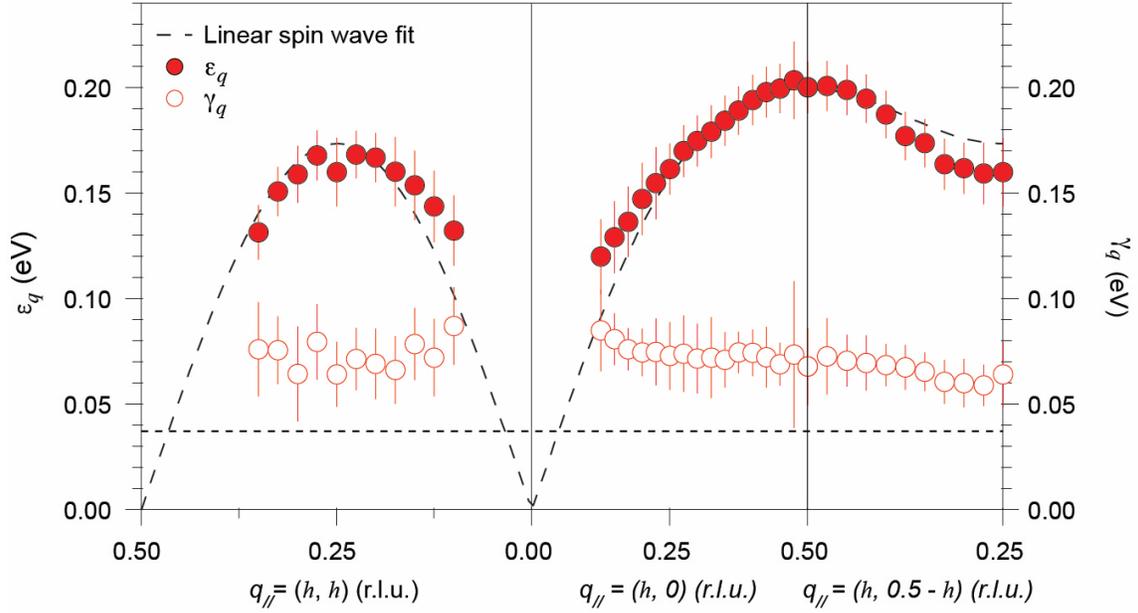

**Fig. 3. Dispersion of magnetic excitations in NdNiO$_2$ and fit to the linear spin wave theory.** A summary of fitted magnetic mode energy $\varepsilon_q$ (filled red circles) and damping factor $\gamma_q$ (empty red circles) versus projected in-plane momentum transfers $q_{//}$ along high-symmetry directions. The dashed curve is linear spin wave dispersion for a two-dimensional antiferromagnetic Heisenberg model fit to the data with $J_1 = 64.1 \pm 3.4$ meV and $J_2 = -10.2 \pm 2.4$ meV. The energy resolution of our RIXS measurement is ~ 37 meV, as indicated in the horizontal dashed line. Error bars of $\varepsilon_q$ are estimated by combining the uncertainty of zero energy-loss position, high energy background, and the standard deviation of the fits. Error bars of $\gamma_q$ are estimated by combining the standard deviation of the fits and the uncertainty of high energy background.

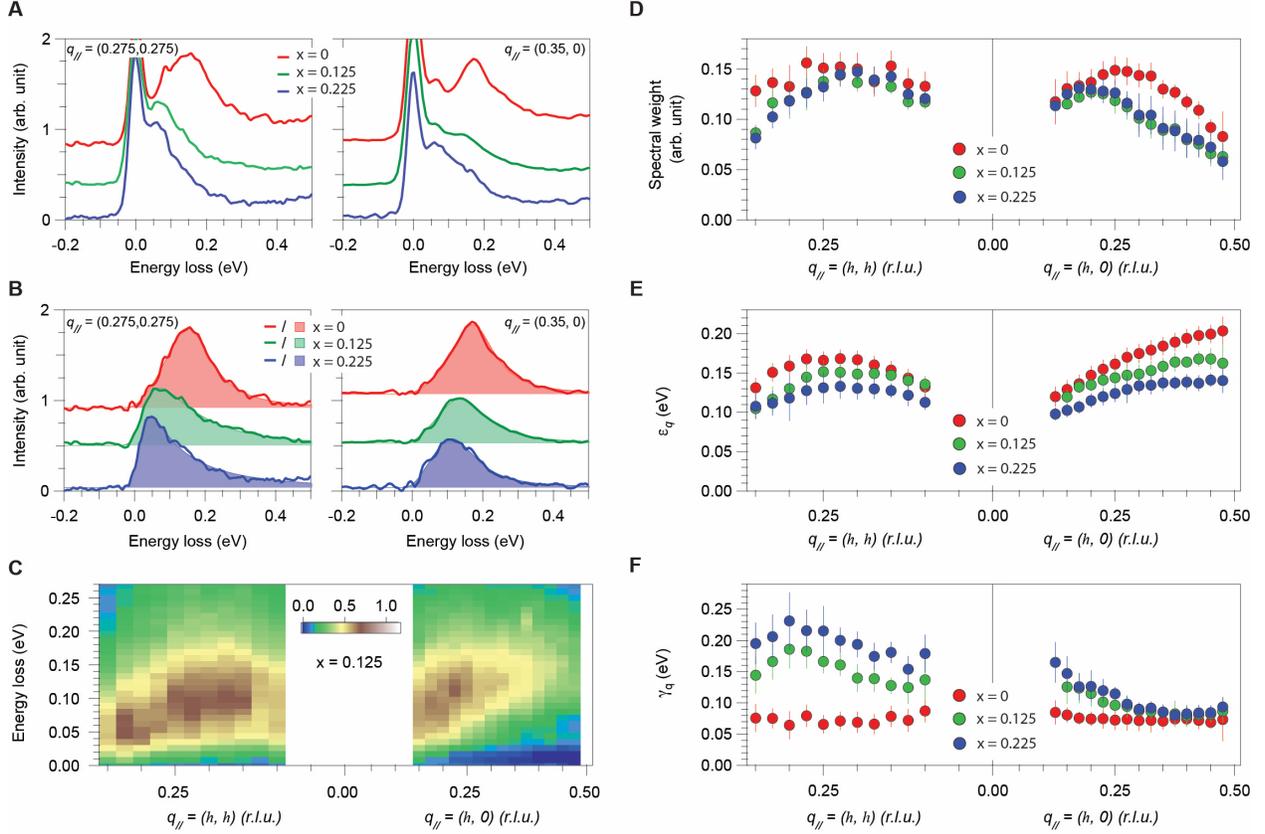

**Fig. 4. RIXS spectra of superconducting Nd$_{1-x}$Sr$_x$NiO$_2$.** (**A**) Raw spectra at representative momentum positions for x = 0, 0.125, and 0.225. (**B**) Magnetic spectra obtained by subtracting the elastic line, phonon, and background from the raw data shown in (A). The shaded areas indicate the associated DHO fitting. (**C**) Magnetic spectra map of x = 0.125 along *hh-* and *h-* directions. (**D** to **F**) Summary of spectral weight (D), mode energy $\varepsilon_q$ (E), and damping factor $\gamma_q$ (F) for the three dopings. For x = 0.125 and 0.225, error bars of spectral weight and $\varepsilon_q$ are estimated by changing the phonon intensity by +/-30 %, while those of $\gamma_q$ are the estimation of the standard deviation to the fit and the uncertainty of the high energy background. For x = 0, error bars of the $\varepsilon_q$ and $\gamma_q$ are the same as those defined in Fig. 3, while error bars of the spectral weight are derived from the fittings.

# Supplementary Materials for

# Magnetic excitations in infinite-layer nickelates

H. Lu[1,3], M. Rossi[1], A. Nag[2], M. Osada[3], D. F. Li[1], K. Lee[3], B. Y. Wang[3], M. Garcia-Fernandez[2], S. Agrestini[2], Z. X. Shen[1,3], E. M. Been[1], B. Moritz[1], T. P. Devereaux[1,3,4], J. Zaanen[5], H. Y. Hwang[1,3], Ke-Jin Zhou[2]\*, W. S. Lee[1]\*

[1]Stanford Institute for Materials and Energy Sciences, SLAC National Accelerator Laboratory and Stanford University, 2575 Sand Hill Road, Menlo Park, CA 94025, USA.

[2]Diamond Light Source, Harwell Campus, Didcot OX11 0DE, United Kingdom.

[3]Geballe Laboratory for Advanced Materials, Departments of Physics and Applied Physics, Stanford University, Stanford, CA 94305, USA.

[4]Department of Materials Science & Engineering, Stanford University, Stanford, CA 94305, USA.

[5]Instituut-Lorentz for theoretical Physics, Leiden University, Niels Bohrweg 2, 2333 CA Leiden, The Netherlands.

Correspondence to: : kejin.zhou@diamond.ac.uk and leews@stanford.edu

**This PDF file includes:**

    Materials and Methods
    Figs. S1 to S5

**Materials and Methods**

Materials:

The precursor $Nd_{1-x}Sr_xNiO_3$ (x = 0, 0.125, and 0.225) thin film of ~ 10 nm on $TiO_2$-terminated $SrTiO_3$ (STO) (001) substrates were synthesized using recently established and optimal growth conditions, followed by deposition of a 5 unit-cell STO capping layer (*16*). To obtain a full transformation to the infinite layer phase ($Nd_{1-x}Sr_xNiO_2$), a topochemical reduction process is adopted with reduction temperature $T_r$ of 260 °C and reduction time $t_r$ of 1 – 3 hours. The symmetric and asymmetric XRD scans using Cu K$\alpha_1$ source confirm that the films are single crystalline and single phase. Resistivity measurements were conducted to determine the onset temperature of the superconductivity in the x = 0.225 thin film to be ~11 K. Note that the x = 0.125 compound is not superconducting (*28*).

Ultrahigh energy resolution RIXS measurements

The Ni $L_3$-edge RIXS were performed at I21-RIXS beamline of the Diamond Light Source in the United Kingdom. The combined energy resolution of the beamline and the RIXS spectrometer was determined by the full width half maximum (FWHM) of the elastic scattering from carbon tape, which was well fitted by a Gaussian function. The resolution for x = 0 and 0.225 data was 37 meV, and that of the x = 0.125 data was 41 meV. The scattering geometry is sketched in Fig. S1. The samples were mounted on a 6-axis in-vacuum manipulator and cooled to around 15 K prior the measurements. The RIXS data were collected with the linear horizontal polarization (parallel to the scattering plane, π-polarization) of the incident X-ray beam. The scattering angle of the spectrometer was fixed at 2θ = 154°. For the momentum dependence measurement shown in Figs. 2, 3, and 4, the incident photon energy was tuned to the maximum of the resonance profile of the magnetic excitation feature as shown in Fig. 1A, which is 0.1 eV lower than the maximum of the absorption curve near the Ni $L_3$-edge. Since the magnetic excitations are quasi-2D (Fig. 2C), the data shown in this manuscript are plotted as a function of in-plane projected momentum transfer $q_{//}$, i.e. projection of $q = k_f - k_i$ onto the NiO$_2$ plane. In this manuscript, the momentum is denoted in reciprocal lattice unit ($2\pi/a$, $2\pi/b$), and the (0, 0) - (0.5, 0) direction is parallel to the Ni–O bond direction. Note that in our convention, positive $q_{//}$ corresponds to grazing-emission geometry, while negative $q_{//}$ corresponds to grazing-incidence geometry. According to polarization-resolved RIXS studies on cuprates (*19*), our scattering geometry yields magnetic signal dominant spectrum at positive $q_{//}$ branch.

Data analysis and fitting

All RIXS data were normalized to incident photon flux. For each momentum position, the elastic (zero energy loss) peak positions were determined by the non-resonant signal from a carbon tape placed near the sample surface, and fine adjusted by the fitted elastic peak position. The fitting model involves a Gaussian for the elastic peak, an anti-symmetrized Lorentzian for the phonon peak, a damped harmonic oscillator function for the magnetic excitation peak, and a tail of an anti-symmetrized Lorentzian for high energy charge background. The model is convolved with a Gaussian function with the FWHM corresponding to the total energy resolution of the RIXS instrument and is then fitted to the data. A representative fitting of the RIXS spectra is shown in Fig. S2A. All data and the associated fits of the undoped compound are shown in Fig. S2B.

For the undoped compound, we have also fitted the magnetic excitation using an anti-Lorentzian function for magnetic excitations, yielding a very similar result for the dispersion relation fitted by the DHO, as shown in Fig. S3.

For the x = 0.125 data, at large momentum positions, phonon is separated from the magnetic excitation peak in the raw data, which can be fitted using the same fitting procedure used for the data of x=0 sample. For these momentum positions, we found that the fitted phonon component is essentially the same as those of the undoped data, as shown in Fig. S4C. This suggests that the phonon component is essentially doping independent. In fact, RIXS phonon studies on cuprates also indicate that the phonon intensity varies ~ 10 % from under- to over-doped regimes in the momentum positions not affected by the charge density wave excitations (*32*). Therefore, for the data at smaller momentum where phonon and magnetic excitations are significantly overlapped in the raw data, we assume the phonon components are identical to those fitted from the undoped data in the fitting process. The same assumption and fitting procedure are also applied for the analysis of x = 0.225 data. Examples at representative momentum positions are demonstrated in Fig. S4D. The error bar of the spectral weight and the mode energy is conservatively estimated by varying the phonon intensity by +/- 30%, substantially larger than the usual variation seen in the case of cuprates.

Linear spin waves

The single spin-flip excitation dispersion was fitted by linear spin wave theory on a square-lattice spin-1/2 AFM Heisenberg model (*25*). We consider the following Heisenberg Hamiltonian

$$H = J_1 \sum_{\langle i,j \rangle} S_i \cdot S_j + J_2 \sum_{\langle i,i' \rangle} S_i \cdot S_{i'}$$

where $J_1$ and $J_2$ are the first- and second-order nearest-neighbor magnetic exchange couplings parameters. Within linear spin wave theory, high-order magnetic exchange couplings beyond second-order could not be determined independently from the data without additional constraints (*25*). Using linear spin wave theory, the dispersion relation is

$$\omega_m = 2Z_C \sqrt{A_q^2 - B_q^2}$$

where

$$A_q = J_1 - J_2[1 - \cos(2\pi h)\cos(2\pi k)]$$

$$B_q = J_1[\cos(2\pi h) + \cos(2\pi k)]/2$$

We fixed the renormalization factor $Z_-$ at 1.18, which is expected for a 2D spin-1/2 square-lattice Heisenberg AFM (*25*).

Spectral Moment Analysis

Although the resolution of RIXS is relatively poor compared to neutron scattering, it collects the average properties of the imaginary part of the dynamical spin susceptibility over the full energy scale associated with the spin-spin interactions and the damping, asserting that the separation of scales characteristic of a doped Mott insulator is in effect. In fact, the success of the DHO fitting form Eq. 1 implies that the experiment reveals the lowest three spectral moments of the magnetic excitation spectra. The spectral moments are defined as $A_n(q) = \int d\omega \omega^n A(q,\omega)$ (*33*); it is easy to check that for a bosonic susceptibility the overall spectral weight is associated with the zero-th moment $A_0(q)$, while the average energy and damping are set by the first- and second moment of the spectral distribution: $A_1(q)$ and $\sqrt{A_2(q)}$, respectively. In turn, these spectral moments are determined by the quantum time evolution involving short, microscopic times. For arbitrary Hamiltonians these are computable (*33*) through the "operator spreading" identifications $A_0(k) = \langle |\vec{M}_k|^2 \rangle$, $A_1(k) = \langle |\vec{M}_k[H, \vec{M}_k]| \rangle$, $A_2(k) = \langle |\vec{M}_k[H,[H, \vec{M}_k]]| \rangle$, ... where $\vec{M}_k$ is the magnetization operator at wave vector $k$.

In a doped Mott insulator, the system of localized spins responsible for magnetism does not change at short times, other than the carriers corresponding to missing spins, which delocalize as described by the t-J model (*31*). In simple t-J model the spectral weight $A_0(k)$ decreases slowly with doping concentration as some spins are replaced by holes. The first moment $A_1(q)$ softens compared to magnons at half-filling, as every hole removes 4 exchange bonds. Corrections to the exchange terms, or additional exchange pathways, due to the doped holes mitigate this softening (*29, 30*). Vigorous hole motion will diminish the spin life times, and increase the second moment, such that one expects overdamped, relaxational dynamics. When this can be captured by an energy independent damping parameter the spectra do not contain more information and higher moments can be ignored. As noted previously and even more appropriate in doped systems, when the spectrum is overdamped the positive- and negative frequencies overlap, and one has to use the damped harmonic oscillator function (Eq. 1) leading to strongly asymmetric line shapes when damping $\gamma_q$ is larger than the mode energy $\varepsilon_q$.

Fig. S5 show the first three spectral moments, and most significantly the overall spectral weight (Fig. S5B) is only mildly reduced, consistent with spin dilution. *Indeed the superconducting compound acts as a doped Mott-insulator,* with an average energy (*i.e.* the first moment $A_1(q)$) only slightly softened compared to the magnons in the parent compound, and a similar dispersion (Fig. S5C). The most significant change is in the second moment, which increases dramatically by up to a factor of three compared to the parent nickelate (Fig. S5D). Note that the spectral moment analysis agrees well with the mode energy $\varepsilon_q$ and damping $\gamma_q$ obtained from damped harmonic oscillator function fitting shown in the main text, namely $A_1(q) \approx \varepsilon_q$ and $\sqrt{A_2(q)} \approx \gamma_q$ (compare Figs. 4D-F and Figs. S5B-D).

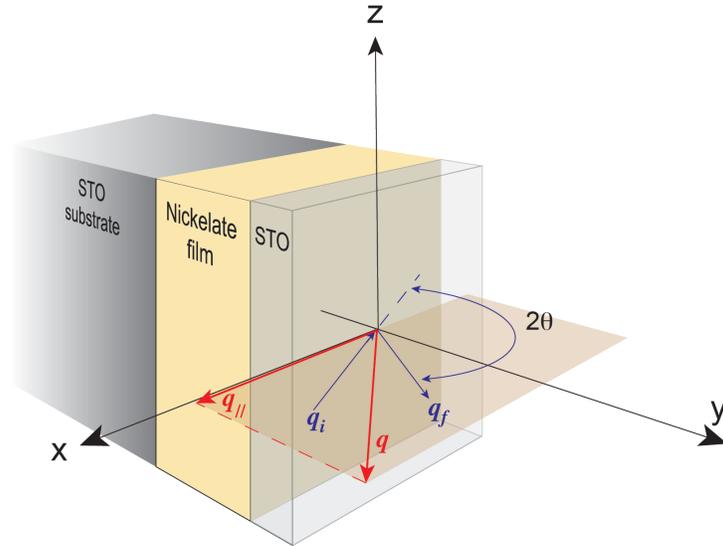

**Fig. S1. Scattering geometry of the RIXS experiment.** The infinite-layer nickelate films were fabricated on a SrTiO$_3$ (STO) substrate. A STO capping layer is grown on the nickelate film to support its crystalline order. The scattering angle 2θ is fixed at 154°. The y-axis is always along the (001) direction of the film. The sample is rotated with respect to the axis z, which is perpendicular to the scattering plane (light brown plane). $q_i$, $q_f$, and $q$ represent the momentum of incident photon, scattering photon, and the momentum transfer $q = q_f - q_i$. Sample rotation effectively changes the projected in-plane momentum transfer, $q_{//}$. The sample can also rotate with respect to the y-axis to access different high symmetry directions in reciprocal space.

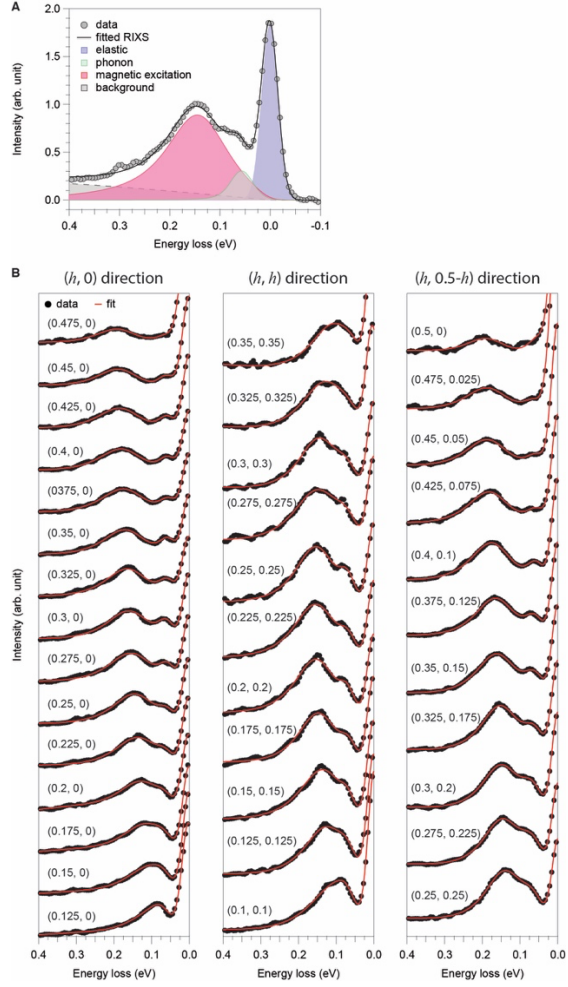

**Fig. S2. Fitting to RIXS spectrum of NdNiO$_2$.** (**A**) A representative example of the fit. The fit function includes a Gaussian for the elastic line (blue shaded region), anti-symmetrized Lorentzian for the phonon (green), damped harmonic oscillator for the magnetic excitation (red), and an anti-symmetrized Lorentzian for the high energy background (grey). Each fit function component is convolved with a Gaussian function with a FWHM of 37 meV to capture the energy resolution of the RIXS measurements and then fit to the data. (**B**) Raw data of the undoped parent compound NdNiO$_2$ with the fits (red curves) superimposed. The corresponding projected in-plane momentum is indicated in parenthesis for each spectrum.

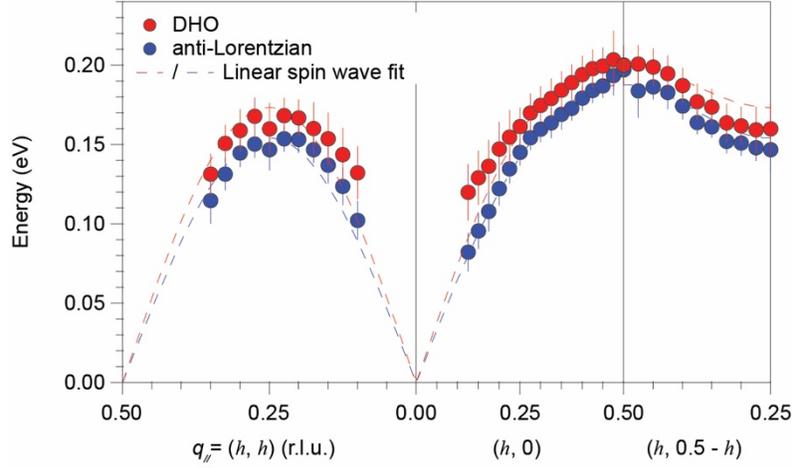

**Fig. S3. Dispersion of magnetic excitations in NdNiO$_2$ fitted by damped harmonic oscillator and anti-Lorentzian functions.** The dispersion of magnetic excitations obtained by fitting the magnetic peak to a damped harmonic oscillator (DHO, red) and an anti-symmetrized Lorentzian (blue) function. Both show similar results. The dashed curves are linear spin wave dispersion for a two-dimensional antiferromagnetic Heisenberg model fitted to the respective dispersion data. For the damped harmonic oscillator fit, $J_1$ = 64.1 ± 3.4 meV, $J_2$ = -10.2 ± 2.4 meV, and renormalization factor $Z_c$ = 1.183 ± 0.014, while the anti-symmetrized Lorentzian fit, $J_1$ = 51.1 ± 1.8 meV, $J_2$ = -14.2 ± 1.3 meV, and the renormalization factor $Z_c$ = 1.183 ± 0.008.

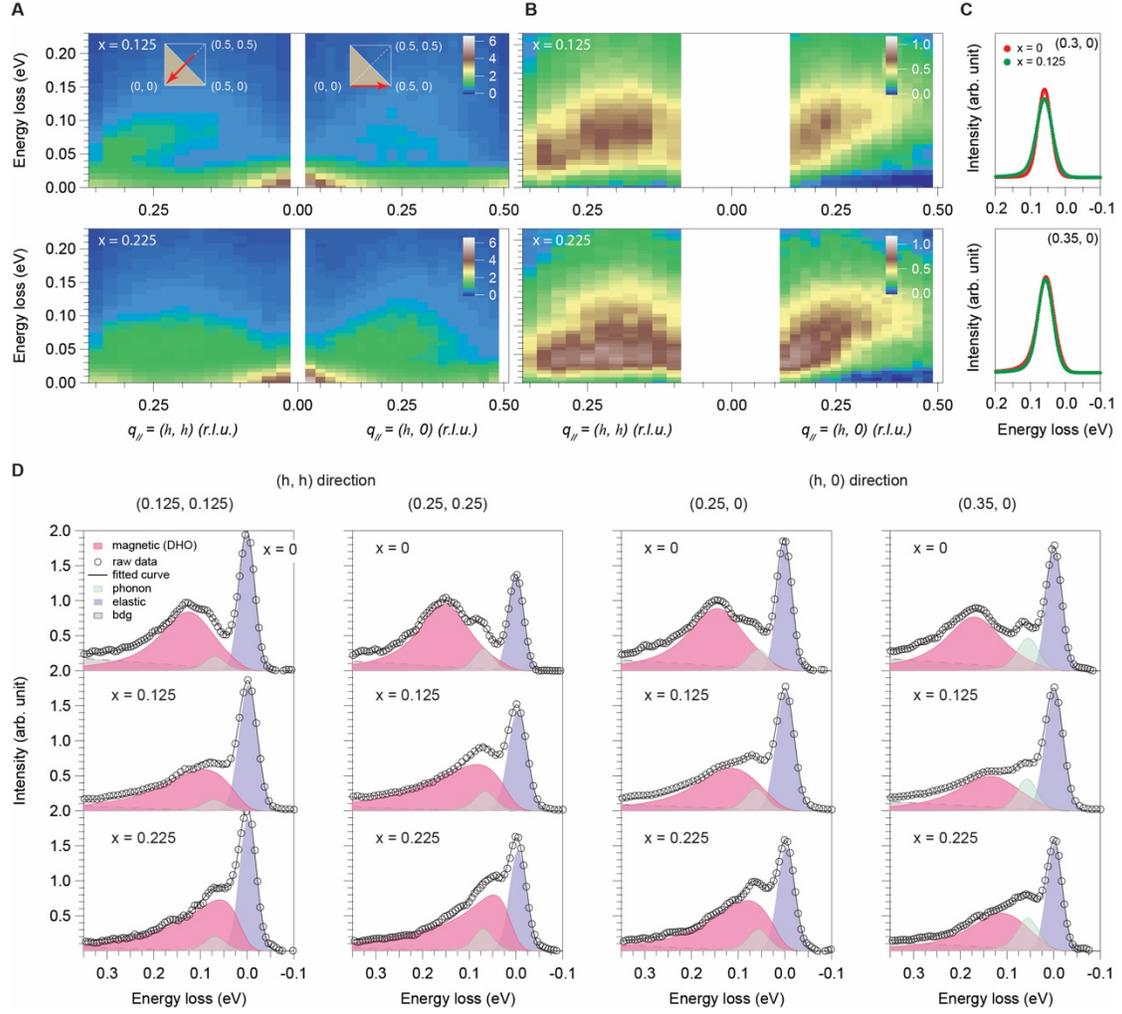

**Fig. S4. Raw RIXS map of Nd$_{1-x}$Sr$_x$NiO$_2$ and Fitting analysis at representative momentum positions**. (**A**) Raw RIXS intensity maps along *hh-* and *h-* directions for x = 0.125 and 0.225. (**B**) Magnetic spectra maps of x = 0.125 and 0.225 along *hh-* and *h-* directions, which were obtained by subtracting the fitted elastic peak, phonons, and background components from the raw data. (**C**) The fitted Phonon component of x = 0 and 0.125 data at representative large momentum, exhibiting little doping dependence. (**D**) The upper row shows data of the parent compound NdNiO$_2$ with the same fitting described in the Methods and Fig. S2. The middle and lower row show data of the x = 0.125 and 0.225 compound with corresponding fit by assuming the same phonon component in the undoped parent compounds except for the (0.35, 0) data of the x = 0.125 compound.

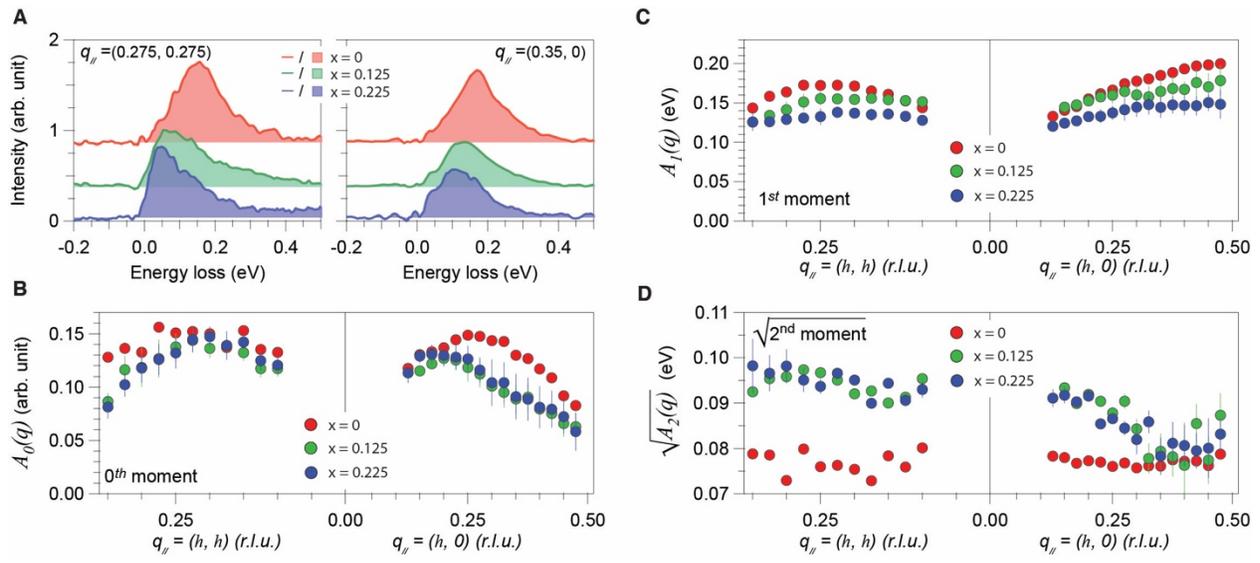

**Fig. S5. Spectral momentum analysis for $Nd_{1-x}Sr_xNiO_2$.** (**A**) Magnetic spectra obtained by subtracting the elastic line, phonon, and background from the raw data. (**B** to **D**) Summary of the zeroth (B), the first (C), and the square root of the second spectral moments (D) for x = 0, 0.125, and 0.225. For x = 0.125 and 0.225 compounds, error bars of zeroth, first, and second moments are estimated by varying the phonon intensity by +/-30 %.